\documentclass{article}
\pdfoutput=1 
\usepackage{nips07submit_e,times}
\usepackage{graphicx}
\usepackage{float}
\usepackage{array}
\usepackage{amsmath}
\usepackage{amssymb}
\newcommand{\bx} {{\mathbf x}}

\newcommand{\bX} {{\mathbf X}}

\newcommand{\bfi} {{\mathbf \phi}}
\DeclareMathSymbol{\R}{\mathalpha}{AMSb}{"52}
\newcommand{\barre}{\rule[5mm]{\textwidth}{0.5pt}\\~\vspace*{-5mm}~}%
\newcommand{\SF}{\bf \sffamily}

\title{Locality and low-dimensions in the prediction of natural experience from fMRI
\thanks{Advances in Neural Information Processing Systems 20,
  Sch\"olkopf B., Platt J. and Hofmann T. (Editors), \copyright \; MIT Press, 2008}}
\author{
Fran\c{c}ois G. Meyer\\
Center for the Study of Brain, Mind and Behavior, \\
Program in Applied and Computational Mathematics\\
Princeton University\\
\texttt{fmeyer@colorado.edu}
\And
Greg J. Stephens\\
Center for the Study of Brain, Mind and Behavior, \\
Department of Physics\\
Princeton University\\
\texttt{gstephen@princeton.edu}
\And
\\
Both authors contributed equally to this work.
}

\begin{document}
\maketitle
\begin{abstract}
  Functional Magnetic Resonance Imaging (fMRI) provides dynamical
  access into the complex functioning of the human brain, detailing
  the hemodynamic activity of thousands of voxels during hundreds of
  sequential time points. One approach towards illuminating the
  connection between fMRI and cognitive function is through decoding;
  how do the time series of voxel activities combine to provide
  information about internal and external experience? Here we seek
  models of fMRI decoding which are balanced between the simplicity of
  their interpretation and the effectiveness of their prediction.  We
  use signals from a subject immersed in virtual reality to compare
  global and local methods of prediction applying both linear and
  nonlinear techniques of dimensionality reduction.  We find that the
  prediction of complex stimuli is remarkably low-dimensional,
  saturating with less than 100 features.  In particular, we build
  effective models based on the decorrelated components of cognitive
  activity in the classically-defined Brodmann areas. For some of the
  stimuli, the top predictive areas were surprisingly transparent,
  including Wernicke's area for verbal instructions, visual cortex for
  facial and body features, and visual-temporal regions for
  velocity. Direct sensory experience resulted in the most robust
  predictions, with the highest correlation ($c \sim 0.8$) between the
  predicted and experienced time series of verbal instructions.
  Techniques based on non-linear dimensionality reduction (Laplacian
  eigenmaps) performed similarly.  The interpretability and relative
  simplicity of our approach provides a conceptual basis upon which to
  build more sophisticated techniques for fMRI decoding and offers a
  window into cognitive function during dynamic, natural experience.
\end{abstract}

\section{Introduction}
\begin{figure}[htb]
\centerline{
  \includegraphics[width=0.9\textwidth]{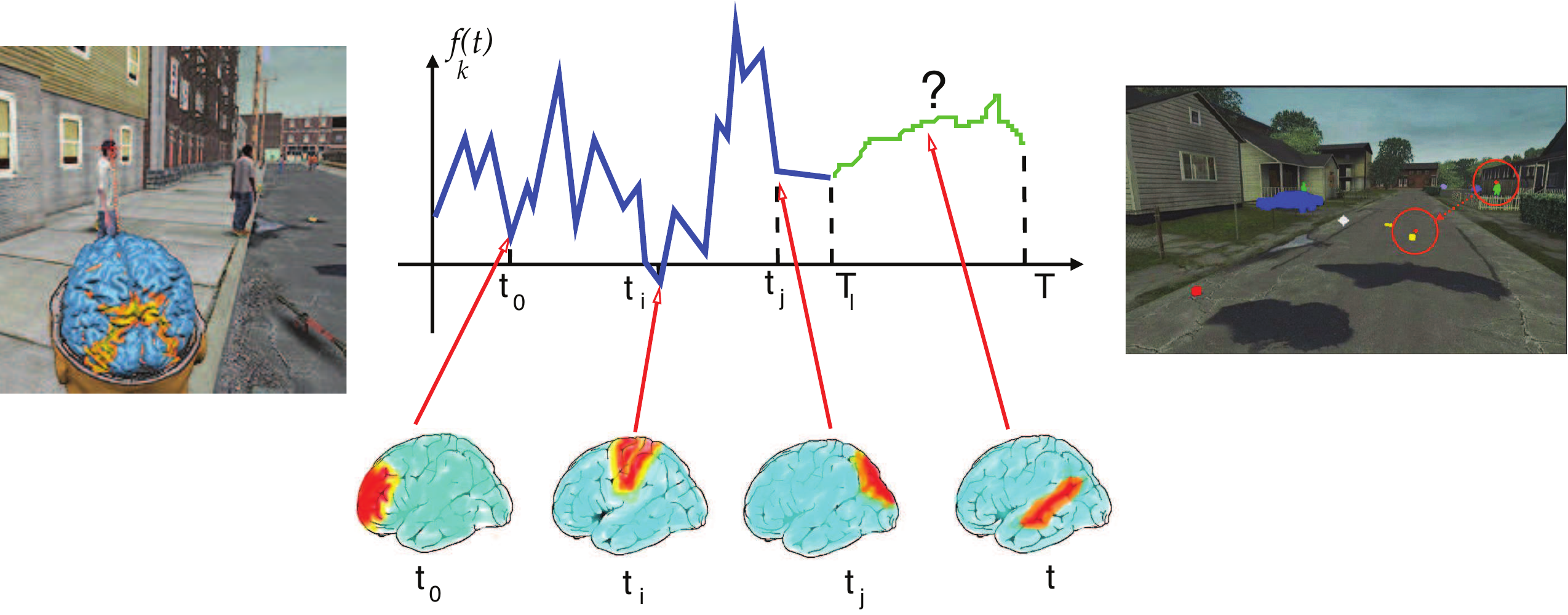}
}
\caption{We study the variation of the set of features $f_k(t), k =1,
  \cdots,  K$ as a function of the dynamical changes in the fMRI signal $\bX(t) =
  [x_1(t), \cdots, x_N(t)]$ during natural experience.  The features represent both external stimuli such as the presence of faces and internal emotional states
  encountered during the exploration of a virtual urban environment (left and right images).
  We predict the feature functions $f_k$ for $t =T_{l+1},\cdots T$, from the knowledge of the entire fMRI dataset $\cal X$, and the
partial knowledge of $f_k(t)$ for $t=1, \cdots, T_l$.
The  ``toy'' activation patterns (middle diagram) illustrate
  the changes in ``brain states'' occurring as a function of time. 
  \label{problem}}
\end{figure}

Functional Magnetic Resonance Imaging \mbox{(fMRI)} is a non-invasive
imaging technique that can quantify changes in cerebral venous oxygen
concentration.  Changes in the fMRI signal that occur during brain
activation are very small (1-5\%) and are often contaminated by noise
(created by the imaging system hardware or physiological processes).
Statistical techniques that handle the stochastic nature of the data
are commonly used for the detection of activated voxels. Traditional
methods of analysis -- which are designed to test the hypothesis that
a simple cognitive or sensory stimulus creates changes in a specific
brain area -- are unable to analyze fMRI datasets collected in
``natural stimuli'' where the subjects are bombarded with a multitude
of uncontrolled stimuli that
cannot always be quantified \cite{Golland07,Malinen07}.\\
The Experience Based Cognition competition (EBC) \cite{EBC} offers an
opportunity to study complex responses to natural environments, and to
test new ideas and new methods for the analysis of fMRI collected in
natural environments.  The EBC competition provides fMRI data of three
human subjects in three 20-minute segments (704 scanned samples in
each segment) in an urban virtual reality environment along with
quantitative time series of natural stimuli or features (25 in total)
ranging from objective features such as the presence of faces to
self-reported, subjective cognitive states such as the experience of
fear.  During each session, subjects were audibly instructed to
complete three search tasks in the environment: looking for weapons
(but not tools) taking pictures of people with piercings (but not
others), or picking up fruits (but not vegetables).  The data was
collected with a 3T EPI scanner and typically consists of the activity
of 35000 volume elements (voxels) within the head.  The feature time
series was provided for only the first two segments (1408 time
samples) and competitive entries are judged on their ability to
predict the feature on the third segment (704 time samples, see
Fig. \ref{problem}).
At a microscopic level, a large number of internal variables associated
with various physical and physiological phenomena contribute to the dynamic
changes in the fMRI signal. Because the fMRI signal is a large scale (as
compared to the scale of neurons) measurement of neuronal activity, we
expect that many of these variables will be coupled resulting in a low
dimensional set for all possible configurations of the activated fMRI
signal. In this work we seek a low dimensional representation of the entire
fMRI dataset that provides a new set of `voxel-free'' coordinates to study
cognitive and sensory features.

We denote a three-dimensional volumes of fMRI composed of a total of $N$
voxels by $\bX (t) = [x_1(t), \cdots, x_N(t)]$.  We have access to $T$ such
volumes. We can stack the spatio-temporal fMRI dataset into a $N \times T$
matrix,
\begin{equation}
{\cal X}=
\begin{bmatrix}
x_1(1) &  \cdots & x_1(T) \\
 \vdots  & \vdots & \vdots  \\
x_N(1) &  \cdots & x_N(T)
\end{bmatrix},
\end{equation}
where each row $n$ represents a time series $\bx_n$ generated from voxel
$n$ and each column $j$ represents a scan acquired at time $t_j$.  We call
the set of features to be predicted $f_k, k=1,,\cdots,K$.  We are
interested in studying the variation of the set of features $f_k(t), k =1,
\cdots, K$ describing the subject experience as a function of the dynamical
changes of the brain, as measured by $\bX(t)$. Formally, we need to build
predictions of $f_k(t)$ for $t = T_{l+1},\cdots T$, from the knowledge of the
entire fMRI dataset $\cal X$, and the partial knowledge of $f_k(t)$ for
the training time samples $t=1, \cdots, T_l$ (see Fig. \ref{problem}).
\begin{figure}[htb]
\centerline{
  \includegraphics[width=0.5\textwidth]{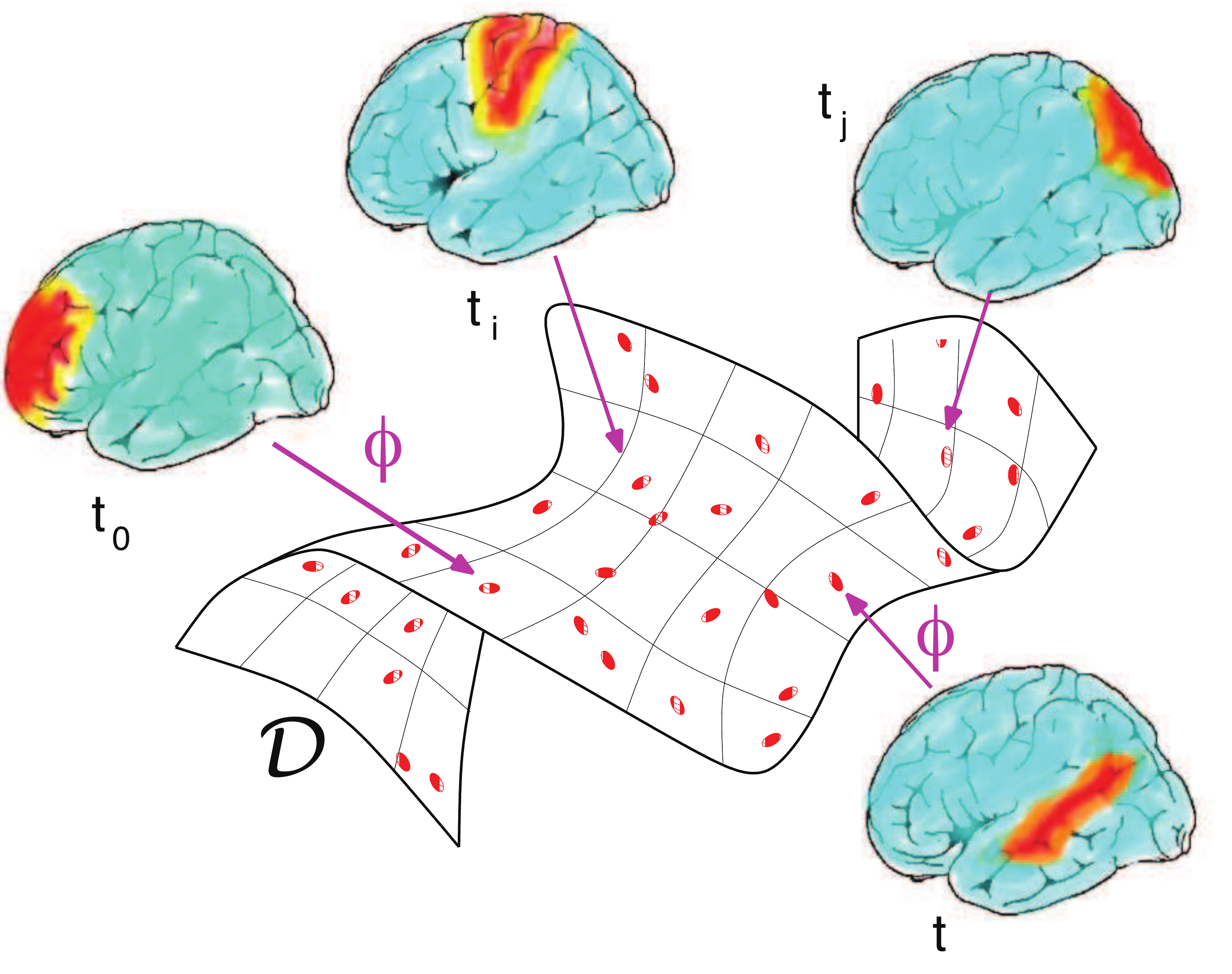}
}
\caption{Low-dimensional parametrization of the set of ``brain
  states''. The parametrization is constructed from the  samples provided
  by the fMRI data at different times, and in different states.
\label{brainstates}}
\end{figure}
\section{A voxel-free parametrization of brain states}
We use here the global information provided by the dynamical evolution of
$\bX(t)$ over time, both during the training times and the test times.  We
would like to effectively replace each fMRI dataset $\bX(t)$ by a small set
of features that facilitates the identification of the brain states, and
make the prediction of the features easier. Formally, our goal is to
construct a map $\phi$ from the voxel space to low dimensional space.
\begin{alignat}{2}
\phi:    \R^N & \mapsto {\cal D} \subset \R^L\\
\bX(t) = [x_1(t),\cdots,x_N(t)]^T & \mapsto (y_1(t),\cdots,y_L(t)),
\end{alignat}
where $L \ll N$.  As $t$ varies over the training and the test sets,
we hope that we explore most of the possible brain configurations that
are useful for predicting the features. The map $\phi$ provides a
parametrization of the brain states. Figure \ref{brainstates} provides
a pictorial rendition of the map $\phi$. The range $\cal D$,
represented in Fig. \ref{brainstates} as a smooth surface, is the set
of parameters $y_1,\cdots,y_L$ that characterize the brain dynamics.
Different values of the parameters produce different ``brain states'',
associated with different patterns of activation. Note that time does
not play any role on $\cal D$, and neighboring points on $\cal D$
correspond to similar brain states.  Equipped with this
re-parametrization of the dataset $\cal X$, the goal is to learn the
evolution of the feature time series as a function of the new
coordinates $[y_1(t),\cdots,y_L(t)]^T$.  Each feature function is an
implicit function of the brain state measured by $
[y_1(t),\cdots,y_L(t)]$.  For a given feature $f_k$, the training data
provide us with samples of $f_k$ at certain locations in $\cal D$. The
map $\phi$ is build by globally computing a new parametrization of the
set $\left \{ \bX(1),\cdots,\bX(T)\right \}$. This parametrization is
built into two stages. First, we construct a graph that is a proxy for
the entire set of fMRI data $\left \{ \bX(1),\cdots,\bX(T)\right
\}$. Second, we compute some eigenfunctions $\phi_k$ defined on the
graph. Each eigenfunctions provides one specific coordinate for each
node of the graph.
\subsection{The graph of brain states}
We represent the fMRI dataset for the training times and test times by
a graph.  Each vertex $i$ corresponds to a time sample $t_i$, and we
compute the distance between two vertices $i$ and $j$ by measuring a
distance between $\bX(t_i)$ and $\bX(t_j)$. Global changes in the
signal due to residual head motion, or global blood flow changes were
removed by computing a a principal components analysis (PCA) of the
dataset ${\cal X}$ and removing a small number components. We then used
the $l^2$ distance between the fMRI volumes (unrolled as $N \times 1$
vectors).  This distance compares all the voxels (white and gray
matter, as well as CSF) inside the brain.
\subsection{Embedding of the dataset}
Once the network of connected brain states is created, we need a
distance to distinguish between strongly connected states (the two
fMRI data are in the same cognitive state) and weakly connected states
(the fMRI data are similar, but do not correspond to the same brain
states).  The Euclidean distance used to construct the graph is only
useful locally: we can use it to compare brain states that are very
similar, but is unfortunately very sensitive to short-circuits created
by the noise in the data. A standard alternative to the geodesic
(shortest distance) is provided by the average commute time,
$\kappa(i,j)$, that quantifies the expected path length between $i$
and $j$ for a random walk started at $i$. Formally, $\kappa(i,j) =
H(j,i) + H(i,j)$, where $H(i,j)$ is the hitting time,
\begin{equation*}
H(i,j) = E_i[T_j] \quad \text{with} \quad T_j=\min \{n\ge 0; Z_n =j\},
\end{equation*}
for a random walk $Z_n$ on the graph with transition probability
$\mathbf P$, defined by $P_{i,j} = w_{i,j}/d_i$, and $d_i = D_{i,i} =
\sum_j w_{i,j}$ is the degree of the vertex $i$.  The commute time can
be conveniently computed from the eigenfunctions $\phi_1, \cdots,
\phi_N$ of ${\mathbf N}= {\mathbf D}^{\frac{1}{2}}{\mathbf P}{\mathbf
  D}^{-\frac{1}{2}}$, with the eigenvalues $-1 \leq \lambda_N \cdots
\leq \lambda_2 < \lambda_1 =1$.  Indeed, we have
\begin{equation*}
\kappa(i,j)=\sum_{k=2}^N \frac{1}{1 -\lambda_k} 
\left (\frac{\phi_k(i)}{\sqrt{d_i}} -\frac{\phi_k(j)}{\sqrt{d_j}} \right)^2.
\end{equation*}
As proposed in \cite{Belkin03,Berard94,Coifman06a}, we define an  embedding 
\begin{equation}
i \mapsto I_k(i) = 
\frac{1}{1-\lambda_k} \frac{\phi_k(i)}{\sqrt{d_i}}, \quad k=2,\cdots,N
\label{embed}
\end{equation}
Because $-1 \leq \lambda_N \cdots \leq \lambda_2 < \lambda_1 =1$, we have
$\frac{1}{\sqrt{1-\lambda_2}} > \frac{1}{\sqrt{1-\lambda_3}} > \cdots
\frac{1}{\sqrt{1-\lambda_N}}$. We can therefore neglect
$\frac{\bfi_k(j)}{\sqrt{1-\lambda_k}}$ for large $k$, and reduce the
dimensionality of the embedding by using only the first $K$ coordinates in
(\ref{embed}). The spectral gap measures the difference between the first
two eigenvalues, $\lambda_1 -\lambda_2 = 1 -\lambda_2$. A large spectral
gap indicates that the low dimensional will provide a good approximation.
The algorithm for the construction of the embedding is summarized in
Fig. \ref{algo1}.
\begin{figure}[H]
{\SF Algorithm 1: Construction of the embedding}\\
  \barre
  \begin{itemize}
  \item [] {\SF Input}: 
    \begin{itemize}
    \item $\bX (t), t=1,\cdots,T$, $K$: number of eigenfunctions.
    \end{itemize}
  \item []{\SF Algorithm:}
    \begin{enumerate}
    \item construct the graph defined by the $n_n$ nearest neighbors
    \item find the first $K$ eigenfunctions, $\phi_k$, of $\mathbf N$
    \end{enumerate}
  \item  {\SF Output:} For $t_i= 1:T$ 
    \begin{itemize}
    \item  new co-ordinates of $\bX(t_i)$: 
        $y_k(t_i) = \frac{1}{\sqrt{\pi_i}} \frac{\bfi_k(i)}{\sqrt{1-\lambda_k}}
        \quad k=2,\cdots,K+1$
    \end{itemize}
  \end{itemize}
  \barre
  \caption{Construction of the embedding
    \label{algo1}}
\end{figure}
A parameter of the embedding (Fig. \ref{algo1}) is $K$, the number of
coordinates.  $K$ can be optimized by minimizing the prediction
error. We expect that for small values of $K$ the embedding will not
describe the data with enough precision, and the prediction will be
inaccurate. If $K$ is too large, some of the new coordinates will be
describing the noise in the dataset, and the algorithm will overfit
the training data.  Fig. \ref{fig:errDim}-(a) illustrates the effect
of $K$ on the performance of the nonlinear dimension reduction. The
quality of the prediction for the features: faces, instruction and
velocity is plotted against $K$. Instructions elicits a strong
response in the auditory cortex that can be decoded with as few as 20
coordinates. Faces requires more (about 50) dimensions to become
optimal. As expected the performance eventually drops when additional
coordinates are used to describe variability that is not related to the
features to be decoded. This confirms our hypothesis that we can
replace about 15,000 voxels with 50 appropriately chosen coordinates.
\subsection{Semi-supervised learning of the features}
The problem of predicting a feature $f_k$ at an unknown time $t_u$ is
formulated as kernel ridge regression problem. The training
set $\{f_k(t)$ for $t=1, \cdots, T_l\}$  is used to estimate the optimal
choice of weights in the following model,
\begin{equation*}
\hat{f}(t_u) = \sum_{t=1}^{T_l} \hat{\alpha} (t) {\cal K}(y(t_u),y(t)),
\end{equation*}
where $\cal K$ is a kernel and $t_u$ is a time point where we need to predict.
\subsection{Results}
We compared the nonlinear embedding approach (referred to as global
Laplacian) to dimension reduction obtained with a PCA of $\cal X$.
Here the principal components are principal volumes, and for each time
$t$ we can expand $\bX(t)$ onto the principal components.

The 1408 training data were divided into two subsets of $704$ time
samples. We use $f_k(t)$ in a subset to predict $f_k(t)$ in the other
subset. In order to quantify the stability of the prediction we randomly
selected 85 \% of the training set (first subset), and predicted 85 \%
of the testing set (other subset). The role, training or testing, of each subset of
$704$ time samples was also chosen randomly. We generated 20
experiments for each value of $K$, the number of predictors.  The
performance was quantified with the normalized correlation between the
model prediction and the real value of $f_k$,
\begin{equation}
  r=\langle \delta f_k^{est}(t), \delta f_k(t) \rangle/ \sqrt{\langle
    \delta (f_k^{est})^2 \rangle \langle \delta f_k^2\rangle}, 
\label{corr}
 \end{equation}
 where $\delta f_k=f_k(t)-\langle f_k \rangle$. Finally, $r$ was
 averaged over the 20 experiments. Fig. \ref{fig:errDim}-(a) and (b)
 show the performance of the nonlinear method and linear method as a
 function of $K$. The approach based on the nonlinear embedding yields
 very stable results, with low variance. For both global methods the
 optimal performance is reached with less than 50
 coordinates. Fig. \ref{fig:compareAll} shows the correlation coefficients
 for 11 features, using $K=33$ coordinates. For most features, the
 nonlinear embedding performed better than global PCA.
\section{From global to local
\label{local}}
While models based on global features leverage predictive components from
across the brain, cognitive function is often localized within specific
regions. Here we explore whether simple models based on classical Brodmann
regions provide an effective decoder of natural experience.  The Brodmann
areas were defined almost a century ago (see e.g \cite{Kandel00}) and
divide the cortex into approximately 50 regions, based on the structure and
arrangement of neurons within each region.  While the areas are
characterized structurally many also have distinct functional roles and we
use these roles to provide useful interpretations of our predictive models.
Though the partitioning of cortical regions remains an open and challenging
problem, the Brodmann areas represent a transparent compromise between
dimensionality, function and structure.
  
Using data supplied by the competition, we warp each brain into standard
Talairach coordinates and locate the Brodmann area corresponding to each
voxel.  Within each Brodmann region, differing in size from tens to
thousands of elements, we build the covariance matrix of voxel time series
using all three virtual reality episodes.  We then project the voxel time
series onto the eigenvectors of the covariance matrix (principal
components) and build a simple, linear stimulus decoding model using the
top $n$ modes ranked by their eigenvalues,
  \begin{equation}
  f_k^{est}(t)=\sum_{i=1}^n w_i^k m_i^k(t).
  \label{eq:linearPCA}
 \end{equation}
 where $k$ indexes the different Brodmann areas, $\{w_i^k\}$ are the linear
 weights and $\{m_i^k(t)\}$ are the mode time series in each region. The
 weights are chosen to minimize the RMS error on the training set and have
 a particularly simple form here as the modes are decorrelated,
 $w_i^k=\langle S(t) m_i^k(t) \rangle$.   Performance is measured as
 the normalized correlation $r$  
 (Eq. \ref{corr}) between the model prediction and%

\begin{figure}[H]
\begin{center}
 \includegraphics[width=0.95\textwidth]{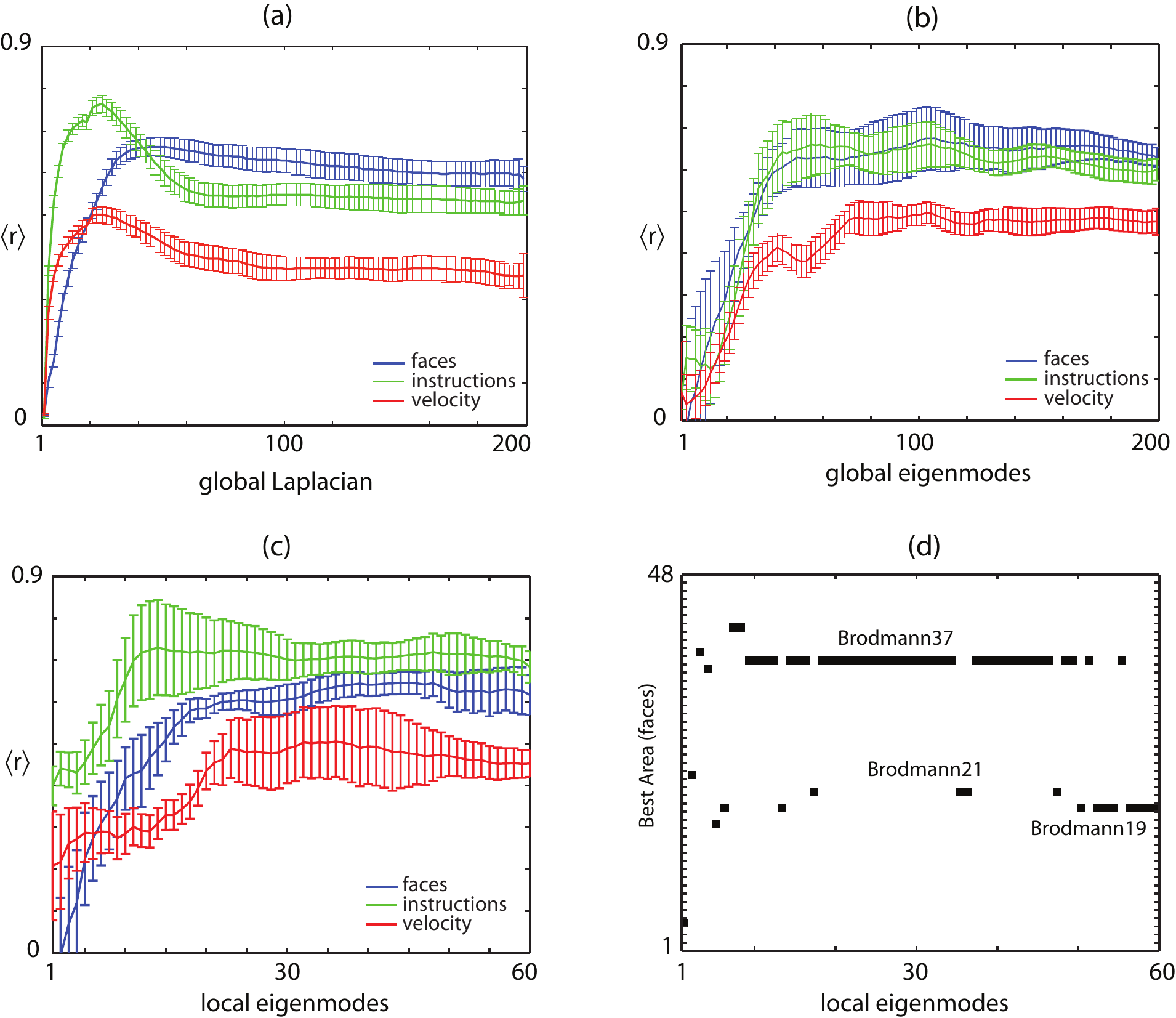}
 \end{center}
\caption{Performance of the prediction of natural experience for three
  features, faces, instructions and velocity as a function of the
  model dimension. (a) nonlinear embedding, (b) global principal
  components, (c) local (Brodmann area) principal components.  In all
  cases we find that the prediction is  remarkably low-dimensional,
  saturating with less than 100 features. (d) stability and
  interpretability of the optimal Brodmann areas used for decoding the
  presence of faces.  All three areas are functionally associated with
  visual processing.   Brodmann area 22 (Wernicke's area) is the best
  predictor of instructions (not shown).  The connections between
  anatomy, function and prediction add an important measure of
  interpretability to our decoding models.
\label{fig:errDim}}
\end{figure}
\noindent  the real stimulus
 averaged over the two virtual reality episodes and we use the region with
 the lowest training error to make the prediction.  In principle, we could
 use a large number of modes to make a prediction with $n$ limited only by
 the number of training samples.  In practice the predictive power of our
 linear model saturates for a remarkably low number of modes in each
 region. In Fig \ref{fig:errDim}(c) we demonstrate the performance of the
 model on the number of local modes for three stimuli that are predicted
 rather well (faces, instructions and velocity).

For many of the well-predicted stimuli, the best
Brodmann areas were also stable across subjects and episodes offering
important interpretability.  For example, in the prediction of instructions
(which the subjects received through headphones), the top region was
Brodmann Area 22, Wernicke's area, which has long been associated with the
processing of human language.  For the prediction of the face stimulus the
best region was usually visual cortex (Brodmann Areas 17 and 19) and for
the prediction of velocity it was Brodmann Area 7, known to be important
for the coordination of visual and motor activity.  Using modes derived
from Laplacian eigenmaps we were also able to predict an emotional state,
the self-reporting of fear and anxiety.  Interestingly, in this case the
best predictions came from higher cognitive areas in frontal cortex,
Brodmann Area 11.

\begin{figure}[htb]
\begin{center}
 \includegraphics[width=0.6\textwidth]{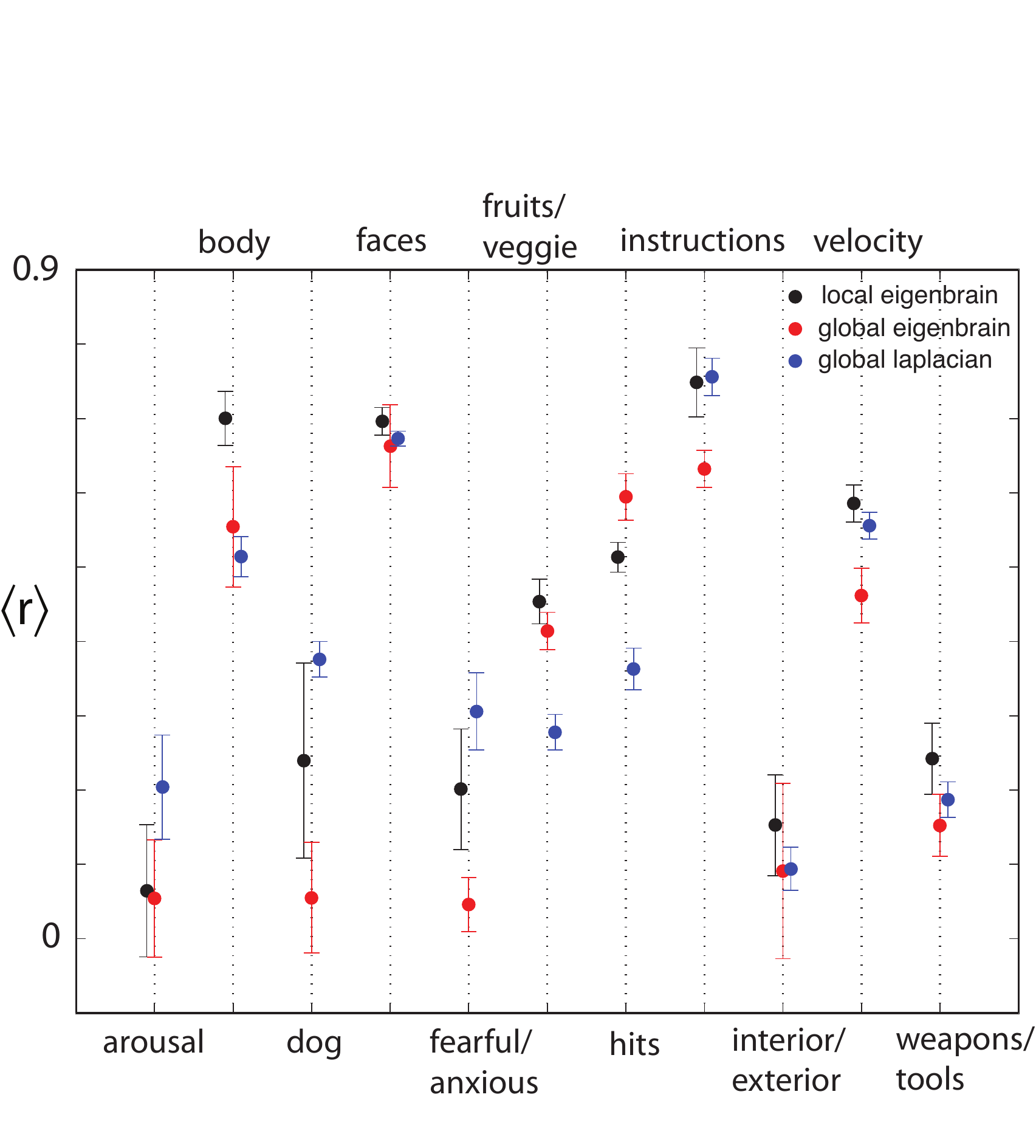}
 \end{center}
 \caption{Performance of the prediction of natural experience for eleven
  features, using three different methods.  Local decoders do well on stimuli
  related to objects while nonlinear global methods better capture stimuli related to emotion.  
 \label{fig:compareAll}}
\end{figure}

While the above discussion highlights the usefulness of classical
anatomical location, many aspects of cognitive experience are not likely to be so
simple.  Given the reasonable results above it's natural to look for
ways of combining the intuition derived from single classical location
with more global methods that are likely to do better in prediction.
As a step in this direction, we modify our model to include multiple
Brodmann areas
\begin{equation}
  f_k^{est}(t)=\sum_{l\in A}  \sum_{i=1}^n w_i^l m_i^l(t),
  \label{eq:combPCA}
 \end{equation}
 where A represents a collection of areas.  To make a prediction using the
 modified model we find the top three Brodmann areas as before (ranked by
 their training correlation with the stimulus) and then incorporate all of
 the modes in these areas (nA in total) in the linear model of Eq
 \ref{eq:combPCA}. The weights $\{w_i^l\}$ are chosen to minimize RMS
 error on the training data. The combined model leverages both the
 interpretive power of single areas and also some of the interactions
 between them.  The results of this combined predictor are shown in Fig.~{\ref{fig:compareAll} (black) and are
 generally significantly better than the single region predictions.  For ease of comparison, we also show the best global results  (both nonlinear Laplacian and global principal components).
 For many (but not all) of the stimuli, the local, low-dimensional linear
model is significantly better than both linear and nonlinear global methods.

\section{Discussion}
Incorporating the knowledge of functional, cortical regions, we used fMRI
to build low-dimensional models of natural experience that performed
surprisingly well at predicting many of the complex stimuli in the EBC
competition.  In addition, the regional basis of our models allows for
transparent cognitive interpretation, such as the emergence of Wernicke's
area for the prediction of auditory instructions in the virtual
environment.  Other well-predicted experiences include the presence of body
parts and faces, both of which were decoded by areas in visual cortex.  In
future work, it will be interesting to examine whether there is a
well-defined cognitive difference between stimuli that can be decoded with
local brain function and those that appear to require more global
techniques.  We also learned in this work that nonlinear methods for
embedding datasets, inspired by manifold learning methods
\cite{Belkin03,Berard94,Coifman06a}, outperform linear techniques in their
ability to capture the complex dynamics of fMRI.
Finally, our particular use of Brodmann areas and linear
methods represent only a first step towards combining prior knowledge of
broad regional brain function with the construction of models for the
decoding of natural experience.  Despite the relative simplicity, an entry
based on this approach scored within the top 5 of the EBC2007 competition
\cite{EBC}.

\section*{Acknowledgments}
GJS was supported in part by National Institutes of Health Grant T32
MH065214 and by the Swartz Foundation.  FGM was partially supported by
the Center for the Study of Brain, Mind and Behavior, Princeton
University. The authors are very grateful to all the members of the
center for their support and insightful discussions.


\end{document}